\documentclass[sort&compress,noshowpacs,preprintnumbers,amsmath,amssymb,aps,floatfix,preprint,showkeys]{revtex4}

\usepackage{graphicx}

\begin{document}

\title{Open and anisotropic soft regions in a model polymer glass} 

\author{Carlo Andrea\@ Massa}
\affiliation{Istituto per i Processi Chimico-Fisici-Consiglio Nazionale delle Ricerche (IPCF-CNR), via G. Moruzzi 1, I-56124 Pisa, Italy}

\author{Francesco\@ Puosi}
\affiliation{Istituto Nazionale di Fisica Nucleare, Largo B. Pontecorvo 3, I-56127 Pisa, Italy}
\affiliation{Dipartimento di Fisica ``Enrico Fermi'', 
	Universit\`a di Pisa, Largo B.\@Pontecorvo 3, I-56127 Pisa, Italy}

\author{Antonio\@ Tripodo}
\affiliation{Dipartimento di Fisica ``Enrico Fermi'', 
	Universit\`a di Pisa, Largo B.\@Pontecorvo 3, I-56127 Pisa, Italy}

\author{Dino\@ Leporini}
\email{dino.leporini@unipi.it}
\affiliation{Dipartimento di Fisica ``Enrico Fermi'', 
	Universit\`a di Pisa, Largo B.\@Pontecorvo 3, I-56127 Pisa, Italy}
\affiliation{Istituto per i Processi Chimico-Fisici-Consiglio Nazionale delle Ricerche (IPCF-CNR), via G. Moruzzi 1, I-56124 Pisa, Italy}

\begin{abstract}

The vibrational dynamics of a model polymer glass is studied by Molecular Dynamics simulations. The focus is on the "soft" monomers with high participation to the lower-frequency vibrational modes contributing to the thermodynamic anomalies of glasses.  To better evidence their role, the threshold to qualify monomers as soft is made severe, allowing for the use of systems with limited size.
A marked tendency of soft monomers to form quasi-local clusters involving up to 15 monomers is evidenced. Each chain contributes to a cluster up to about three monomers and a single cluster involves monomer belonging to about 2-3 chains. Clusters with monomers belonging to a single chain are rare. The {\it open} and {\it tenuous}  character of the clusters is revealed by their fractal dimension $d_f < 2$. The inertia tensor of the soft clusters evidences their {\it strong anisotropy} in shape and remarkable {\it linear} correlation of the two largest eigenvalues.  Owing to the limited size of the system, finite-size effects, as well as dependence of the results on the adopted polymer length, cannot  be ruled out.

\end{abstract}

\maketitle

\section{Introduction}

Specific heat and thermal conductivity of amorphous solids exhibit anomalies with respect to crystals \cite{BinderKob}. Customarily, the difference is ascribed to "soft modes" (SMs), i.e. the low-frequency portion of the vibrational density of states (vDOS) $g(\omega)$.  It was noted already 20 years ago that in glassy materials some low-frequency modes are "quasi-localized" with only few particles effectively participating in a mode \cite{LairdSchoberPRL91,SchoberOligschlegerPRB96}. SMs are involved in a well-known universal feature of amorphous solids, namely the boson peak (BP), a SM excess over the Debye level revealed when plotting the reduced vDOS $g(\omega)/ \omega^2$  \cite{BinderKob}. The BP frequency window corresponds to wavelengths where the homogeneous picture of elastic bodies assumed by the Debye model becomes questionable. Therefore, it is of major interest to investigate the SM spatial extension.
More recently, another source of "excess modes" has been identified in computer simulations of model glasses 
\cite{LernerPRL16,MizunoParticipationRatioPNAS17,ShimadaParticipationRatioPRE18,BouchbinderLernerPRL18,AngelaniParisiRuoccoPNAS18,BerthierSzamelParticipationRatioNatCom19}. It is composed of quasi-localized low-frequency modes with a density obeying $g_{loc}(\omega) \sim \omega^4$. They are observed at frequencies significantly lower than BP and the link between the two phenomena is not immediate \cite{BerthierSzamelParticipationRatioNatCom19}. 

Models for the BP dealt with quasi-local vibrational states due to soft anharmonic potentials \cite{BuchenauLocalModeBPPRB91,ParshinSchoberLocalizedBPPRB03}, local inversion-symmetry breaking \cite{ZacconeBP_PRB16},
phonon-saddle transition in the energy landscape \cite{ParisiBPNature03}, elastic heterogeneities
\cite{SchirmacherDiezemannBPElasticHetPRL98,GotzeBPElasticHetPRE00,ElliottBPVanHovePRL01,BarratElasticHetBosonPeakNonAffinePRL06,RuoccoSchirmacherBPElasticHeterogSciRep13}
and broadening and shift of the lowest van Hove singularity in the corresponding reference crystal \cite{ChumakovBPVanHovePRL11} due to the distribution of force constants
\cite{ShengDOSSpringScience91,SchirmacherDiezemannBPElasticHetPRL98,ElliottBPVanHovePRL01}. However, interest in localized SMs extends beyond relationship to theoretical models and BP. It was suggested that SMs are correlated with irreversible structural relaxation in the supercooled liquid state \cite{Harrowell_NP08}, and that SM spatial distribution is correlated with structural relaxation in glassy polymers \cite{RottlerParticipationRatioPolymerGlassSoftMatter14} as well as rearrangements upon mechanical deformation and plasticity  
\cite{ManningLiuPRL11,RottlerRigglemanPRX14}. 

The present paper investigates the degree of the localization of the SMs in an amorphous system made of a dense assembly of linear polymer chains. To this aim, monomers are classified in terms of their softness, i.e. the degree of participation to the lower-frequency vibrational modes and a fraction of "soft" monomers is selected by setting a suitable (high) threshold. Clusters of soft monomers are identified and characterised in terms of their fractal dimension, anisotropy in shape and contributions provided by the monomers of a single chain and multiple chains.


\section{Methods and simulation}
We study by molecular dynamics (MD) simulations a dense system of coarse-grained linear polymer chains made of ten monomers each, resulting in a total number of monomers $N=500$. Each monomer has mass $m$. Non-adjacent monomers in the same chain or monomers belonging to different chains are defined as "non-bonded" monomers. Non-bonded monomers when placed at mutual distance $r$ interact via a shifted Lennard-Jones (LJ) potential:
	\begin{equation}
	U^{LJ}(r) = {\epsilon} \left [   \left (\frac{\sigma^*}{r}\right )^{12} - 2 \left (\frac{\sigma^*}{r}\right )^{6}\right ] + U_{cut}
	\label{Eq:modifiedLJ},
	\end{equation}
where $\sigma^* = 2^{1/6} \sigma$ is the minimum of the potential,  $U^{LJ}(r = \sigma^*) = -\epsilon + U_{cut}$. The potential is truncated at $r=r_c=2.5\sigma$ for computational convenience and the constant $U_{cut}$ adjusted to ensure that $U^{LJ}(r)$ is continuous at  $r=r_c$ with $U^{LJ}(r)=0$ for $r \ge r_c$.  Adjacent monomers in the same chain are bonded by the harmonic potential ${\mathrm{U^b}\left(r\right)=k\left(r-r_0\right)^2}$; in the following, results from systems with different values of the spring stiffness, $k = 500, 1000,  2500$ in units of $\epsilon/\sigma^2$, are shown. Since no torsional or bending potentials are present, the chain exhibits high flexibility.

All the data presented in the work are expressed in reduced MD units: length in units of $\sigma$, temperature in units of $\epsilon/k_B$, where $k_B$ is the Boltzmann constant, and time in units of $\tau_{MD} = (m\sigma^2 / \epsilon)^{1/2}$. We set $\sigma = 1$, $\epsilon = 1$,  $m = 1$ and $k_B = 1$  \cite{UnivPhilMag11}. 

Simulations were carried out with the open-source Molecular-Dynamics (MD) software LAMMPS \cite{PlimptonLAMMPS,PlimptonURL}.  The system was initially equilibrated at temperature $T=1.25$ and pressure $p=4.7$, then cooled with the same pressure at $T=0.7$ and finally quenched to $T=0.001$ with pressure $p=0$ in a single time step equal to 0.0002. A subsequent waiting time of $200$ time units was allowed to relax the system. A total number of $154$ amorphous replicas were investigated.

\section{Vibrational modes}
\label{vib}
We consider a solid in which $N$ particles with equal mass $m$  are regarded as point masses free to vibrate with small amplitude  ${\bf u}_i$ about their equilibrium positions ${\bf r}_i$  (i= 1,2, ...,N ) and let the total potential energy be denoted as  $U({\bf r}_1, \dots, {\bf r}_N)$ \cite{Bell_1972,ZacconeBosonPeakPolymerMM18}. In the harmonic approximation the equation of motion can  be written in terms of the  Hessian ${\bf H}$ of the system:
\begin{equation}
m \ddot{{\bf u}} = - {\bf H} {\bf u}
\label{EqMot}
\end{equation}
where ${\bf u}$ is the displacement field, ${\bf u} = ({\bf u}_1, \dots, {\bf u}_N)$. The elements of the Hessian are defined as second derivatives of the potential energy of the system under mechanical equilibrium:
\begin{equation}
	H_{ij}^{\alpha\beta}=\dfrac{\partial^2U}{\partial x_{i,\alpha}\partial x_{j,\beta}}
	\label{equ:Hessian}
\end{equation}
where $x_{i, \alpha}$ ($\alpha =1,2,3$) are three-dimensional Cartesian components of the displacements of the $i$-th monomer. We can convert Eq.\ref{EqMot} into an eigenvalue problem by performing a time Fourier transform, which gives
\begin{equation}
m \omega_l^2 \hat{{\bf u}}_l =  {\bf H} \; \hat{{\bf u}}_l
\label{equ:EigenValues}
\end{equation}
where $\omega_l$ is the $l$-th eigenfrequency of the system ($l=1,\cdots, 3N$, with $\omega_m > \omega_n$ if $m>n$) and $ \hat{{\bf u}}_l $ is the corresponding eigenvector (displacement field) with normalization
\begin{equation}
\sum_{i, \alpha} \hat{u}_{i,\alpha,l} \hat{u}_{i,\alpha,l'} = \delta_{l,l'}
\label{norm}
\end{equation}
The {\it participation fraction} of particle $i$ in eigenmode $ \hat{{\bf u}}_l$  is defined by  \cite{Harrowell_NP08,HarrowellJCP09}:
\begin{equation}
p_{i}(\omega_l) = \sum_{\alpha} |  \hat{u}_{i,\alpha,l} |^2
\label{pfDef}
\end{equation}
Eq.\ref{norm} and eq.\ref{pfDef} yield the following relation providing the normalization of the participation fraction:
\begin{equation}
\sum_{i} p_{i}(\omega_l)  = 1
\label{pfNorm}
\end{equation}
A useful metric of the spatial extension of the $l$-th mode is the {\it participation ratio} \cite{Bell_1972,HarrowellJCP09,RottlerParticipationRatioPolymerGlassSoftMatter14,MizunoParticipationRatioPNAS17,ShimadaParticipationRatioPRE18,BerthierSzamelParticipationRatioNatCom19}
\begin{equation}
P(\omega_l) = \left[ N \sum_{i} p^2_{i}(\omega_l)  \right] ^{-1}
\label{prNorm}
\end{equation}
If the mode is completely delocalized so that all particles contribute equally, $p_{i}(\omega_l) \sim 1/N$ and $P(\omega_l) = 1$. Instead, a mode localized on a single particle $i_0$ leads to $p_{i}(\omega_l) = \delta_{i,i_0}$ and $P(\omega_l) = 1/N$. For a plane wave, $P(\omega_l) = 2/3$ \cite{HarrowellJCP09,BerthierSzamelParticipationRatioNatCom19}.

Finally, in order to quantify the softness of a particle, we consider the overall participation fraction of the $i$-th particle to the first $N_m$  modes and define the softness field as \cite{RottlerParticipationRatioPolymerGlassSoftMatter14} :
\begin{equation}
\phi_i = \frac{1}{N_m} \sum_{l=1}^{N_m} p_{i}(\omega_l) 
\end{equation}
We choose $N_m=30$ \cite{Harrowell_NP08}. Therefore, the $i$-th monomer is considered {\it softer} than the $j$-th one if $\phi_i > \phi_j$. 

\begin{figure}[t]
\centering
	\includegraphics[width=8.5 cm]{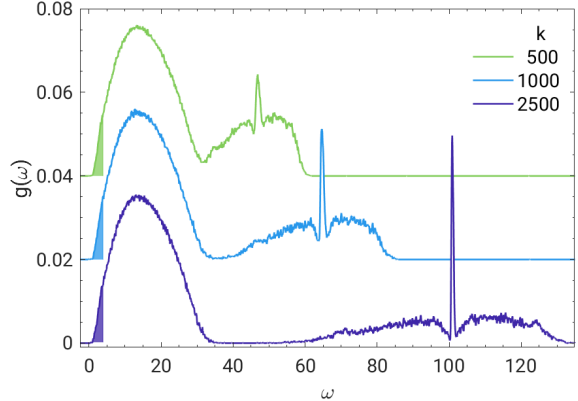}
	\caption{\label{fig:DOS} Vibrational density of states (vDOS) for glassy decamers with different bond  strengths. The thirty lower frequency modes are highlighted. They cover the BP region observed for the present model at $\omega \sim 2$ \cite{GiuntoliLeporiniBPPRL18}. Curves with $k= 500,1000$ are vertically shifted upwards for clarity reasons. }
\end{figure}

\section{Results and discussion}

\subsection{Vibrational density of states}

We have evaluated the vibrational density of states $g(\omega)$:
\begin{equation}
g(\omega) = \frac{1}{3 N-3} \sum_{l=1}^{3N-3} \delta(\omega - \omega_l)
\label{vDOS}
\end{equation}
Fig.\ref{fig:DOS} plots the vibrational density of states . Two main branches can be distinguished: a high-frequency one governed by the bonding interactions  and a low-frequency one governed by non-bonding LJ interactions \cite{ZacconeBosonPeakPolymerMM18,GiuntoliLeporiniBPPRL18}. It is seen that changing the stiffness of the spring bonding adjacent monomers of the same chain, affects only - as expected -  the high-frequency branch, leaving unaffected the low-frequency one. In accordance with this observation we note that the narrow peak appearing in between the two side lobes of the high-frequency branch is located at the characteristic frequency of the vibration of a dumbell with two monomers coupled by a spring, $(4 k/m)^{1/2}$. To date, the low-frequency branch of vDOS attracted most interest since it is involved in thermodynamic anomalies of amorphous solids \cite{BinderKob}. On the other hand, the high-frequency branch observed in polymeric glasses \cite{ZacconeBosonPeakPolymerMM18,GiuntoliLeporiniBPPRL18} deserves wider attention. As an example, we mention the class of shape memory polymers were the presence of hard and soft domains has been reported \cite{LuHuangShapeMemoryPolymers13,Basfar08}.

\subsection{Localization of the states}

\begin{figure}[t]
\centering
	\includegraphics[width=8.5 cm]{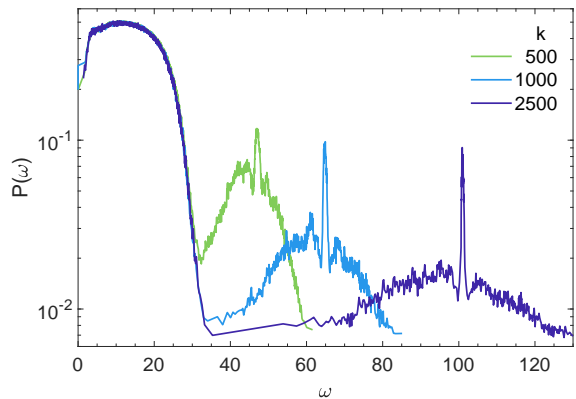}
	\caption{\label{fig:PRtrue} Participation ratio of the mode with frequency $\omega$, Eq.\ref{prNorm}, for different stiffnesses of the spring bonding adjacent monomers of the same chain. For a plane wave, $P(\omega) = 2/3$.}
\end{figure}

Fig.\ref{fig:PRtrue} plots the participation ratio $P(\omega)$, Eq.\ref{prNorm}.  Like vDOS, it shows two branches, a low-frequency one governed by non-bonding LJ interactions ($\omega \lesssim 30$) and a high-frequency branch governed by the bonding interactions. If the bond stiffness is high ($k \gtrsim 1000$), the two branches are well separated. The participation ratio of the low-frequency branch exhibits a maximum at about $0.5$,  close to the one anticipated for the plane waves. The higher localization of the high-frequency modes is explained by noting that the bonding interactions has more local character. It is seen that decreasing the bond strength does not affect the low-frequency branch whereas it increases the participation ratio of the high-frequency branch. The decrease of the participation ratio at very low frequency ($\omega \lesssim 5$), i.e. the higher localization of the softer modes, has been noted in polymers glasses \cite{RottlerSoftMatter10} as well as in atomic glasses 
\cite{MizunoParticipationRatioPNAS17,ShimadaParticipationRatioPRE18,BerthierSzamelParticipationRatioNatCom19}. It will be characterized in the following sections.

\begin{figure}[t!]
\centering
	\includegraphics[width=8.5 cm]{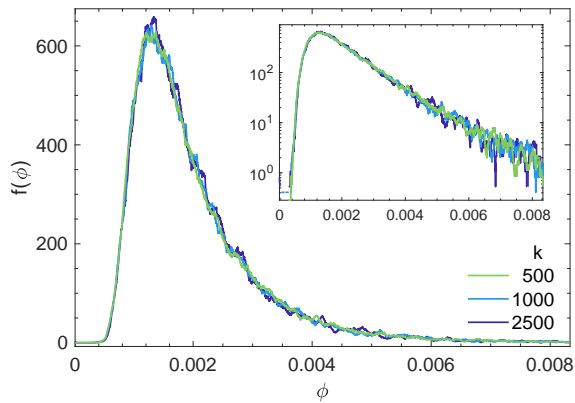}
	\caption{Distribution of the monomer softness $f(\phi)$. Inset: exponential tail of the distribution at large softness.  A monomer is defined to be {\it soft} if $ \phi \ge 4.6 \cdot 10^{-3}$.}
	\label{fig:Pf}
\end{figure}

\subsection{Quasi-local soft regions}

Fig.\ref{fig:Pf} shows the distribution of the particle softness $f(\phi)$. The shape is quite similar to other studies on polymer glasses \cite{RottlerParticipationRatioPolymerGlassSoftMatter14}. It exhibits a nearly exponential tail at high softness. It is seen that the softness is virtually independent of the bond strength. 

\begin{figure}[t]
\centering
	\includegraphics[width=8.5 cm]{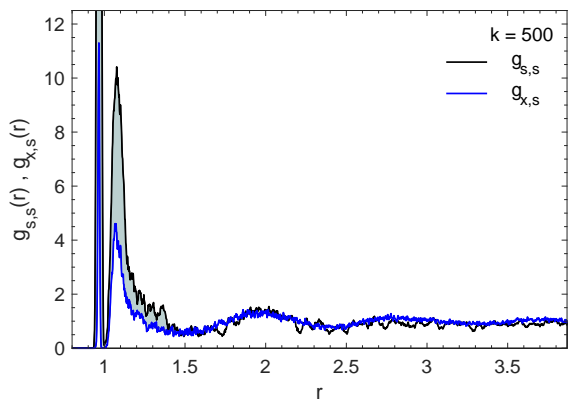}
	\caption{Radial distribution functions $g_{x,s}$ and $g_{s,s}$. $k=500$. The grey region emphasizes the tendency of a central soft monomer to be surrounded by more soft particles than a generic one in the first coordination shell.}
	\label{fig:gXSgSS}
\end{figure}

\begin{figure}[t]
\centering
	\includegraphics[width=8.5 cm]{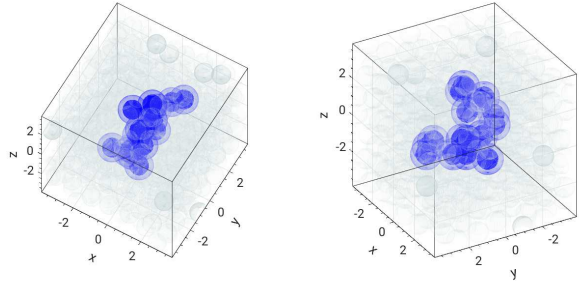}
	\caption{Two views of a snapshot of a typical large soft cluster composed of $15$ monomers. Bond strength $k=10^3$.  Gyration radius $R_g = 1.89$, principal values of the inertia tensor $I_1=   26.4$, $I_2= 30.5$, $I_3=50.4$.  The blue color corresponds to particles with unit diameter (the approximate monomer diameter). Two soft monomers are defined as "close", and then belongs to the cluster, if their surrounding lighter regions superimpose, i.e. their mutual distance $r \le 1.5$, see Fig.\ref{fig:gXSgSS}.}
	\label{visual}
\end{figure}

\begin{figure}[t]
\centering
	\includegraphics[width=14 cm]{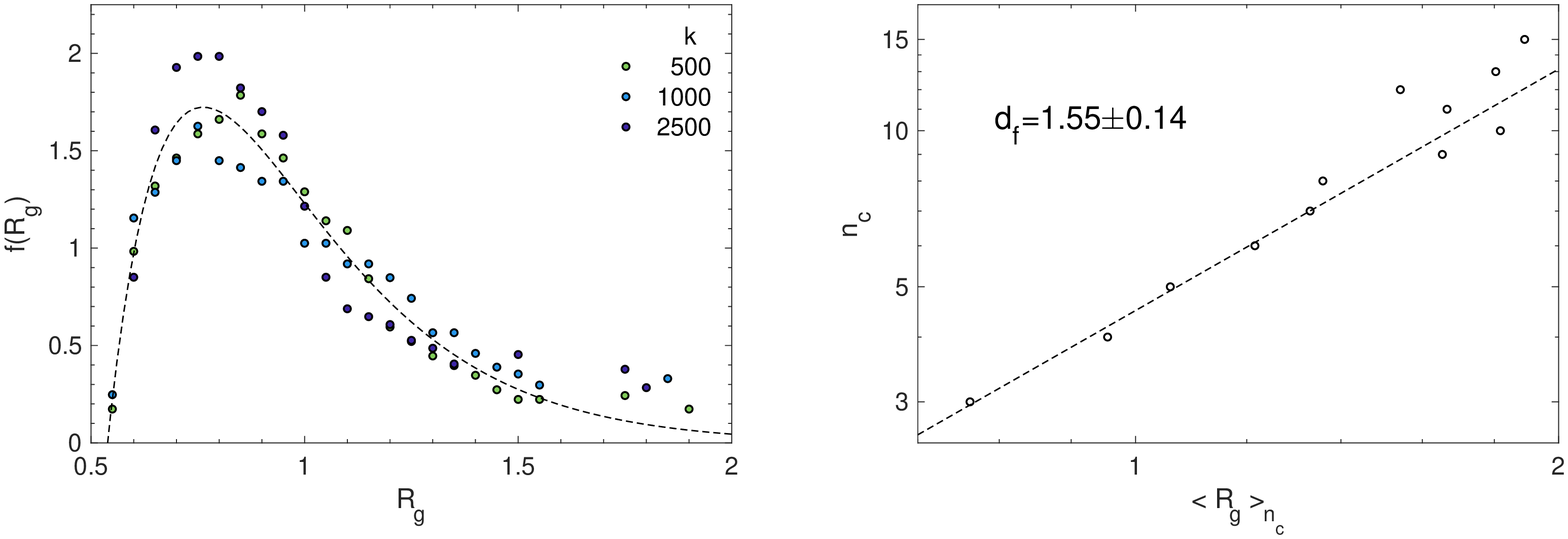}
	\caption{Left: distribution of the radius of gyration of soft clusters. The dashed line is a guide for the eyes. Right: correlation plot between the number of members of a cluster $n_c$ and the gyration radius averaged over all the clusters with the same number of members, $\left < R_g \right >_{n_c}$. The dashed line is the best-fit curve with the power-law, Eq.\ref{fractdim}, by adopting a linear least-squares procedure with weights proportional to the number of clusters involved in each average to draw $\left < R_g \right >_{n_c}$. The resulting fractal dimension is $d_f = 1.55 \pm 0.14$. If {\it no} weight is used, one finds $d'_f = 1.7 \pm 0.3$.}
	\label{fig:rogfig5}
\end{figure}

\subsubsection{Evidence of soft clusters}
A remarkable question is whether the soft particles in glasses are isolated or group together and form clusters \cite{LairdSchoberPRL91,SchoberOligschlegerPRB96,RottlerParticipationRatioPolymerGlassSoftMatter14}. Henceforth a {\it soft monomer} is defined as a monomer with $ \phi \ge 4.6 \cdot 10^{-3}$. The definition of the threshold is more stringent of previous studies where the softest particles have $\phi = 2.7 \cdot 10^{-3}$ \cite{Harrowell_NP08}. Fig.\ref{fig:gXSgSS} plots the radial distribution functions of soft monomers surrounding either a central soft one, $g_{s,s}$, or a central generic one, $g_{x,s}$. It is seen that soft particles tend to be surrounded by more soft particles than a generic one, i.e. they tend to form clusters.

It is worthwhile to characterize the soft clusters evidenced by radial distribution functions. To this aim, by definition, two soft monomers are said to be {\it close} to each other if they are spaced by no more than $r_c$.  We choose $r_c=1.5$, corresponding roughly to the first minimum of $g_{s,s}$ and $g_{x,s}$ according to Fig.\ref{fig:gXSgSS}. A soft cluster of $n_c$ members (with $n_c \ge 3$) is defined as the largest group of soft monomers where each member is close to at least another member. Usually, in a configuration one finds up to three clusters. Fig.\ref{visual} visualises a typical large soft cluster.

\subsubsection{Size and shape of the soft clusters}
In order to characterize the size and the shape of the soft clusters we consider their inertia tensor $\mathbf{ I}$ with respect to the center of mass and evaluate the eigenvalues $I_1, I_2, I_3$ with $I_1<I_2<I_3$. The size of the cluster is estimated by the radius of gyration which is evaluated as 
\begin{equation}
R_g =   \left [ \frac{1}{2 \, m \, n_c} ( I_1+I_2+ I_3 ) \right ]^{1/2}
\label{rgeigenvalues}
\end{equation}
A transparent interpretation of the radius of gyration is given by the usual definition
\begin{equation}
R_g = \left [ \frac{1}{n_c} \sum_{i} \left (r_i^{(CM)} \right )^2 \right ] ^{1/2}
\label{rg}
\end{equation}
where $r_i^{(CM)}$ is the distance of the i-th particle of the soft cluster from the centre of mass of the latter. Fig.\ref{fig:rogfig5}(left) plots the distribution of the radius of gyration. It is roughly as large as about one diameter. We are interested in the fractal dimension of the clusters $d_f$ drawn by the radius of gyration \cite{JungblutFractalDimPCCP19}. Fig.\ref{fig:rogfig5}(right) presents the correlation plot between the number of members of a cluster $n_c$ and the gyration radius averaged over all the clusters with the same number of members $\left < R_g \right >_{n_c}$. The fractal dimension $d_f$ is drawn by best-fitting the data with the power-law:
\begin{equation}
n_c = A \left [ \left < R_g \right >_{n_c} \right ]^{d_f}
\label{fractdim}
\end{equation}
where $A$ is a constant. We follow two different approaches for the best-fit procedure. In one case, the least-squares are weighted with the number of clusters involved in the average $\left < R_g \right >_{n_c}$. This leads to  $d_f = 1.55 \pm 0.14$. On the other hand,  with {\it no} weight, the fit procedure yields $d'_f = 1.7 \pm 0.3$. The fractal dimension points to  soft cluster which are open and tenuous  \cite{JungblutFractalDimPCCP19}. Indeed, the largest identified soft cluster exhibits a loose structure, see Fig.\ref{visual}.

\begin{figure}[t]
\centering
	\includegraphics[width=7 cm]{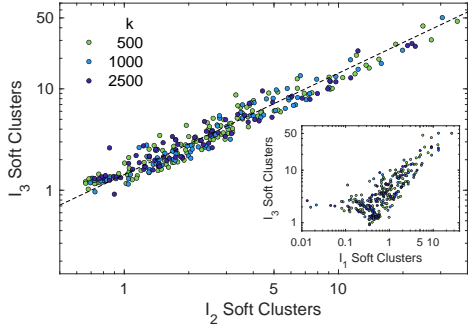}
	\caption{Correlation between the two largest eigenvalues of the inertia tensor. The dashed curve is the best-fit with the straight line  I$_3=\alpha\cdot$I$_2$, $\alpha=1.43\pm0.03$ (Pearson correlation coefficient $r=0.97$). Inset: correlation plot of the largest and the smallest eigenvalues of the inertia tensor.}
	\label{fig:iisuijfig6}
\end{figure}

To provide insight into the shape of the cluster we present in Fig.\ref{fig:iisuijfig6} the correlation plots between the two largest eigenvalues. Strikingly, we find an excellent {\it linear} correlation over more than one decade. Poorer correlation is found between the largest and the smallest eigenvalues, Fig.\ref{fig:iisuijfig6}(inset). The analysis suggests that the soft clusters are {\it anisotropic}  in shape.

\begin{figure}[t]
\centering
	\includegraphics[width=8.5 cm]{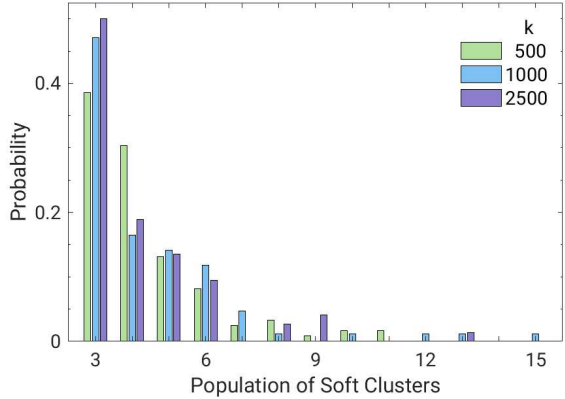}
	\caption{Distribution of the number of soft monomers belonging to a soft cluster.}
	\label{fig:nocfig4}
\end{figure}

\begin{figure}[t]
\centering
	\includegraphics[width=14 cm]{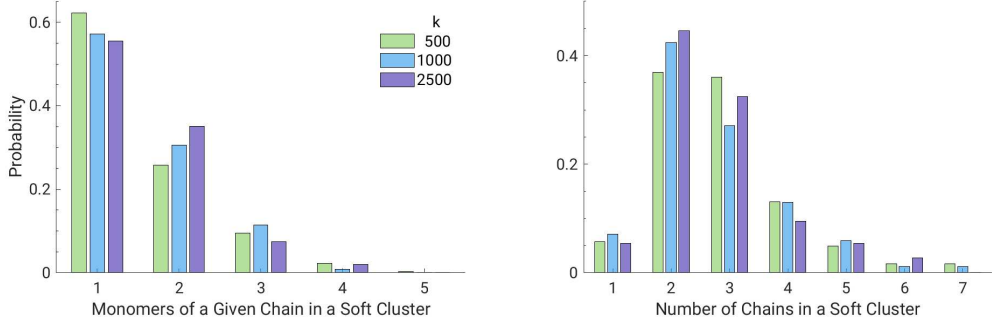}
	\caption{Left: probability of having $n$ monomers belonging to a given chain in a soft cluster.  Right: probability of having monomers coming from $m$ different chains in a given soft cluster.}
	\label{fig:chainclusterfig8}
\end{figure}

\subsubsection{Monomer number and chain partners of the soft clusters}

Fig.\ref{fig:nocfig4} shows the distribution of the number of soft monomers forming a soft cluster. It is seen that the bond strength has only marginal impact on the cluster population. However, there are hints that a stiffer spring favours the formation of soft small clusters.

Finally, Fig.\ref{fig:chainclusterfig8} analyses the relevance of the single chain contribution to a given cluster  and the role of different chains in the formation of a single cluster. Even in this case the dependence on the strength of the bonding interaction is not apparent. Fig.\ref{fig:chainclusterfig8}(left) shows that about up to three soft monomers of a given cluster belong to the same chain. Interestingly, Fig.\ref{fig:chainclusterfig8}(right) evidences that a single cluster is rarely populated by monomers of a single chain, being the most frequent occurrence the involvement of 2-3 different chains.

\section{Conclusions}
Amorphous solids exhibit thermodynamic anomalies rooted in the low-frequency portion of vDOS where SMs are found.
The paper reports on a MD study of the localisation and the shape of SMs in a model polymer glass made of linear chains. Three different variants of the model are considered, having different bonding strength between adjacent monomers of the same chain.  Monomers are classified in terms of softness, i.e. their  participation to the thirty vibrational modes with lowest frequency. The focus is on the fraction of monomers with {\it higher softness} with respect to previous studies, thus resulting in
{\it smaller} collections of particles, justifting the use of limited system sizes.
 Evidence that soft monomers manifest clear tendency to group together in clusters is collected by investigating their radial distribution function, the gyration radius of the clusters as well as their inertia tensor. The study offers two major results, namely the open and tenuous  character of the soft clusters which exhibit a fractal dimension $d_f < 2$ and their anisotropy in shape. A remarkable  {\it linear} correlation of the two largest eigenvalues of the inertia tensor is observed. Owing to the limited size of the system under study, finite-size effects, as well as dependence of the results on the adopted polymer length, cannot  be ruled out. They will be explored in detail in future studies.
\vspace{6pt} 



\begin{acknowledgements}
A generous grant of computing time from Green Data Center of the University of Pisa, and Dell EMC${}^\circledR$ Italia is also gratefully acknowledged. This research was funded by University of Pisa grant number PRA-2018-34 ("ANISE").
\end{acknowledgements}



%






\bibliography{biblio}

%



\end{document}